\documentclass[pra,aps,twocolumn,showpacs,superscriptaddress,floatfix]{revtex4-2}

\usepackage{mathtools}
\usepackage[usenames,dvipsnames]{xcolor}
\usepackage{amsmath}
\usepackage{amssymb,mathrsfs}
\usepackage{graphicx}
\usepackage[colorlinks=true]{hyperref}
\usepackage{soul}
\usepackage{xcolor}
\usepackage{physics}

\usepackage{float}

\newcommand{\prt}{\partial}

\newcommand{\om}{\omega}

\newcommand{\bu}{\mathbf{u}}
\newcommand{\bk}{\mathbf{k}}

\begin{document}
	
	\title{Supersonic Flow Past an Obstacle in a Quasi-Two-Dimensional Lee–Huang–Yang Quantum Fluid}
    
	\author{G. H. dos Santos}
    \affiliation{Instituto de F\'{i}sica, Universidade de S\~{a}o Paulo,
05508-090 S\~{a}o Paulo, SP, Brazil}	

    \author{L. F. Calazans de Brito}
	\affiliation{Instituto de F\'{i}sica, Universidade de S\~{a}o Paulo,
05508-090 S\~{a}o Paulo, SP, Brazil}

	\author{A. Gammal}
  \affiliation{Instituto de F\'{i}sica, Universidade de S\~{a}o Paulo,
05508-090 S\~{a}o Paulo, SP, Brazil}
	
	\author{A. M. Kamchatnov}
	\affiliation{Institute of Spectroscopy, Russian Academy of Sciences, Troitsk, Moscow Region, 108840, Russia}
        \affiliation{Skolkovo Institute of Science and Technology, Skolkovo, Moscow, 143026, Russia}
        \affiliation{ Higher School of Economics, Physical Department, 20 Myasnitskaya ulica, Moscow 101000, Russia}

\begin{abstract}
A supersonic flow past an obstacle can generate a rich variety of wave excitations. This paper investigates, both analytically and numerically, two types of excitations generated by the flow of a Lee–Huang–Yang quantum fluid past an obstacle: linear radiation and oblique dark solitons. We show that wave crests of linear radiation can be accurately described by the proper modification of the Kelvin original theory, while the oblique dark soliton solution is obtained analytically by transformation of the 1D soliton solution to the obstacle's reference frame. A comparison between analytical predictions and numerical simulations demonstrates good agreement, validating our theoretical approach.
\end{abstract}
	
\maketitle
	
\section{Introduction}
	
	
The study of Bose-Einstein condensate~(BEC) has attracted a significant scientific interest due to its quantum fluid behavior. In the condensate of weakly interacting atoms, most problems can be described within the mean-field approximation by the Gross-Pitaevskii equation (GPE)~\cite{pitaevskii16}. However, when the mean-field interspecies attraction in Bose-Bose mixtures exceeds the average repulsive forces, the GPE predicts a collapse of the system. In the case of quantum droplets, this issue can be resolved by incorporating quantum fluctuation effects, leading to a correction of the GPE with  a quartic nonlinear term---the Lee-Huang-Yang (LHY) correction---thus opening the study of the beyond-mean-field effects~\cite{lhy57,petrov15} (see also~\cite{lhrpb21,bulgac02}).
	
By manipulating the interaction strength and the number of atoms, it is possible to create a carefully designed mixture of a two-component BEC in which the cubic mean-field term vanishes, allowing quantum fluctuations to be studied through the quartic nonlinear term~\cite{njj18}. This leads to the LHY equation in the form
\begin{equation}\label{eq1}
    i\hbar \frac{\partial \Psi}{\partial t} = -\frac{\hbar^2 }{2m}\nabla^2 \Psi + U_{ext}\Psi + g_{LHY}|\Psi|^3\Psi,
\end{equation}
where \( U_{ext} \) represents the external potential, which can account for effects such as traps and obstacles, among others. The parameters $g_{LHY}$ and $m$ denote the three-dimensional inter-particle interaction constant and the atomic mass, respectively. The quantum fluid governed by the LHY equation exhibits both nonlinear and dispersive effects. These properties enable the propagation of various atomic excitations, including solitons, vortices, dispersive shock waves, and linear radiation. As in the case of the GPE, a supersonic flow past an obstacle in the LHY fluid is expected to generate stationary coherent wave patterns satisfying the Mach-Cherenkov-Landau resonant radiation condition. These wakes are referred to as Kelvin-like wakes due to their similarity to Lord Kelvin's water wave structure generated by a moving ship (commonly known as the Kelvin wake pattern)~\cite{kelvin-87,whitham-75,johnson-97, kcs-22, carusotto13}. More specifically, for supersonic flow velocities, the GPE predicts two distinct wave structures: oblique dark solitons inside the Mach cone and linear radiation outside it. In the case of the GPE, these structures have been extensively studied, experimentally~\cite{cornell05, amo11} and through theoretical investigations, particularly regarding linear radiation~\cite{gegk-07,kg-13} and oblique dark soliton wakes~\cite{ks-09, el06}. However, for Eq.~\eqref{eq1}, 
this problem remains unsolved.
	
In this work, we develop an analytical framework for the investigation of a supersonic flow past an obstacle in a quasi-2D quantum fluid described by Eq.~\eqref{eq1}. We restrict our analysis to flow velocities that support the propagation of two distinct wake patterns: oblique dark solitons and linear radiation. The paper is structured as follows: Section~II presents the analytical theory of linear radiation outside the Mach cone. Section~III focuses on oblique solitons inside the Mach cone. In Section~IV, we compare analytical predictions with numerical simulations. Finally, we conclude with a discussion in Section~V.

\section{Dimensional Reduction and the Quasi-2D Hydrodynamical Model}

In experimental setups, it is not possible to realize truly low-dimensional systems. Instead, an effective low-dimensional regime can be obtained by ``squeezing'' a 3D BEC, tightly confining it in one or two spatial directions using external potentials. Theoretically, beyond-mean-field energy corrections in Bose gases have been investigated within the framework of 3D--1D and 3D--2D dimensional crossovers in Refs.~\cite{mmp24, tobias_etall18, pawell_etall18}.

In our model, we construct a 3D--2D dimensional crossover by imposing strong confinement along the $z$-axis, such that the system's dynamics are effectively restricted to the $x$--$y$ plane. This means that the harmonic potential corresponds to much greater frequency $\om_z$ in $z$ of atomic motion direction than the frequency $\omega_\perp$ of oscillations in the ``pancake'' plane, $\omega_z / \omega_\perp \gg 1$. Consequently, the atoms condense into the ground state $\phi(z)$ of the motions in the $z$-direction for which the weak interatomic interaction can be neglected, whereas the atomic motion in the $(x,y)$-plane is governed by the Gross-Pitaevskii equation with account of the interatomic interaction. If number of atoms $N$ in the condensate is large enough, so that the radius $R_0$ of the pancake-shape density distribution is much greater than the characteristic harmonic length  $a_z = \sqrt{\hbar/m\om_z}$ in the axial direction, then we can factorize the 3D condensate wave function in the following way,
\begin{equation}\label{eq:separada}
    \Psi(x,y,z,t) = \Psi(x,y,t)\,\phi(z),
\end{equation}
where

\begin{equation}\label{eq:diracao_z}
    \phi(z) = \frac{1}{\pi^{1/4} a_z^{1/2}} e^{- \left( \frac{z^2}{2a_z^2} \right)}
\end{equation}
is the ground-state wave function in the axial direction. Substitution of this expression into Eq.~\eqref{eq1} yields after integration over the $z$-direction the quasi-2D LHY quantum fluid equation:
\begin{equation}\label{eq:adm}
    i\hbar \frac{\partial \Psi(\vb{r},t)}{\partial t} = \left(-\frac{\hbar^2}{2m} \nabla_{\perp}^2
    + U_{\perp} + g_{LHY}^{\prime} |\Psi|^3 \right) \Psi(\vb{r},t),
\end{equation}
where we keep the notation $\Psi(\vb{r},t)$ for the effective 2D condensate wave function with $\vb{r}=(x,y)$. In particular, the number of atoms can be estimated as $N \approx |\Psi|^2\pi R_0^2$. $U_{\perp}$ denotes the potential acting in the $(x,y)$-plane, for our purpose it is convenient to take $U_{\perp}=0$, and $g_{LHY}^{\prime}$ is the renormalized LHY interaction constant, which is defined as 
\begin{equation}
    g'_{\text{LHY}} = \sqrt{\frac{2}{5 \pi^{3/2} a_z^3}}  g_{\text{LHY}}.
\end{equation}
Making the transformation
\begin{equation} 
    \vb{r} \rightarrow \xi_0\vb{r} , \quad  t \rightarrow \tau t, \quad \text{and} \quad \Psi \rightarrow \Psi_0 \Psi
\end{equation}
where $\xi_0 = \hbar/\sqrt{m|\Psi_0|^3g'_{LHY}}$, and $\tau = \hbar/|\Psi_0|^3 g'_{LHY}$, we transform Eq.~(\ref{eq:adm}) to non-dimensional variables. As a result, we arrive at the equation
\begin{equation} \label{eq:LHY_adim}
    i\frac{\partial \Psi(x,y,t)}{\partial t} = -\frac{1}{2} \nabla_{\perp}^2 \Psi(x,y,t) + |\Psi|^3 \Psi(x,y,t).
\end{equation}

To obtain basic properties of a uniform BEC we make the so-called Madelung transformation,

\begin{equation} \label{eq3}
	\Psi(\mathbf{r}, t) = \sqrt{\rho(\mathbf{r}, t)}e^{i\int \mathbf{u}(\mathbf{r})\,d\mathbf{r}}.
\end{equation} 
Then Eq.~\eqref{eq3} transforms to the Euler-like hydrodynamic equations:
\begin{align} \label{eq4}
		\frac{\partial \rho}{\partial t} + \nabla \cdot (\rho \mathbf{u}) = 0, \nonumber \\
		\frac{\partial \mathbf{u}}{\partial t} + (\mathbf{u} \cdot \nabla)
\mathbf{u} +\frac{3}{2}\sqrt{\rho} \nabla \rho = \nabla \left[ \frac{\nabla^2 \rho}{4\rho}
- \frac{(\nabla \rho)^2}{8\rho^2} \right],
\end{align}
where, in the context of the Bose gas theory, $\rho$ and $\bu$ represent the density and velocity fields of the hydrodynamic system, respectively. These equations admit a simple solution $\rho=\rho_0=\mathrm{const}$, $\bu=\bu_0=\mathrm{const}$ which represents the flow of BEC with a uniform density and constant velocity. We are interested in wave patterns generated by such a flow past an obstacle which can be modelled by an additional potential $U_{\perp}$ localized around the origin $(x=0,y=0)$ of our coordinate system.

\section{Linear radiation in LHY fluid flowing past an obstacle}

At first, we have to find the dispersion relation for linear waves propagating along the uniform BEC with $\rho=\rho_0=\mathrm{const}$, $\bu=\bu_0=\mathrm{const}$. To perform the linearization, we separate the smooth background parameters from an oscillatory perturbation, that is, we assume
\begin{equation}
    \mathbf{u} = (u_0 + u', v'), \quad \rho = \rho_0 + \rho',
\end{equation}
where $u'$, $v'$, $\rho'$ denote small perturbations and we take the background flow directed along $x$-axis, $\bu=(u_0,0)$. In the first order approximation with respect to small variables, we get from the system \eqref{eq4} the linearized system:
\begin{equation}\label{eq5}
\begin{split}
	&\frac{\prt \rho '}{\prt t} + u_0 \frac{\prt \rho '}{\prt x} + \rho_0 \left( \frac{\prt u '}{\prt x} + \frac{\prt v '}{\prt y} \right) = 0, \\
	&\frac{\prt u '}{\prt t} + u_0\frac{\prt u '}{\prt x} + \frac{3}{2}\sqrt{\rho_0}\frac{\prt \rho '}{\prt x} =\\
	&\hspace{2cm} \frac{1}{4\rho_0} \left( \frac{\prt^3 \rho '}{\prt x^3} +   \frac{\prt^3 \rho ' }{\prt x \prt y^2} \right),  \\
	&\frac{\prt v '}{\prt t} + u_0\frac{\prt v '}{\prt x} + \frac{3}{2}\sqrt{\rho_0}\frac{\prt \rho '}{\prt y} = \\
	&\hspace{2cm} \frac{1}{4\rho_0} \left( \frac{\prt^3 \rho '}{\prt y^3} + \frac{\prt^3 \rho ' }{\prt y \prt x^2} \right).  \\		
\end{split}
\end{equation}
The dispersion relation is obtained by assuming that the perturbation terms take the form of a plane wave solution,
\begin{equation}
   u', v', \rho' \propto \exp \left[i\left(k_x x + k_y y - \omega t\right)\right],
\end{equation}
so the substitution of this {\it ansatz} into Eq.~\eqref{eq5} yields
\begin{equation} \label{eq6}
    \omega = u_0 k_x + k \sqrt{\frac{3}{2} \rho_0^{\frac{3}{2}} + \frac{k^2}{4}}.
\end{equation}
For a fluid at rest with $u_0=0$, this equation is recognized as the Bogoliubov dispersion relation~\cite{bogolyubov-47}. In the long wavelength limit $k\ll1$, this relation reduces to $\om\approx c_0k$, where
\begin{equation}
    c_0 = \frac{\omega}{k} =  \sqrt{\frac{3}{2}}\rho_0^{\frac{3}{4}}
\end{equation}
is called the `sound velocity'. Introduction of this characteristic flow velocity allows one to rewrite the dispersion relation, making the transformation to dimensionless variables
\begin{equation}\label{eq7}
\omega \rightarrow c_0^2 \omega,\quad k \rightarrow c_0 k,\quad \text{ and }\quad u_0 \rightarrow c_0 M,
\end{equation}
so we get
\begin{equation} \label{eq8}
    \omega \equiv G(k_x,k_y) = M k_x + k \sqrt{ 1 + \frac{k^2}{4}},
\end{equation}
where $M = u_0/c_0$ denotes the Mach number and $k$ is the modulus of the wave vector $\mathbf{k}=(k_x, k_y)$. This form of the dispersion relation coincides with that for linear waves in usual BEC, whose dynamics is described by the standard GPE (see, e.g. Ref.~\cite{gegk-07}), so we can use the method of Refs.~\cite{whitham-75,gegk-07} for finding the Kelvin-like wake pattern.
\begin{figure}[t]
    \centering
    \includegraphics[width=6cm]{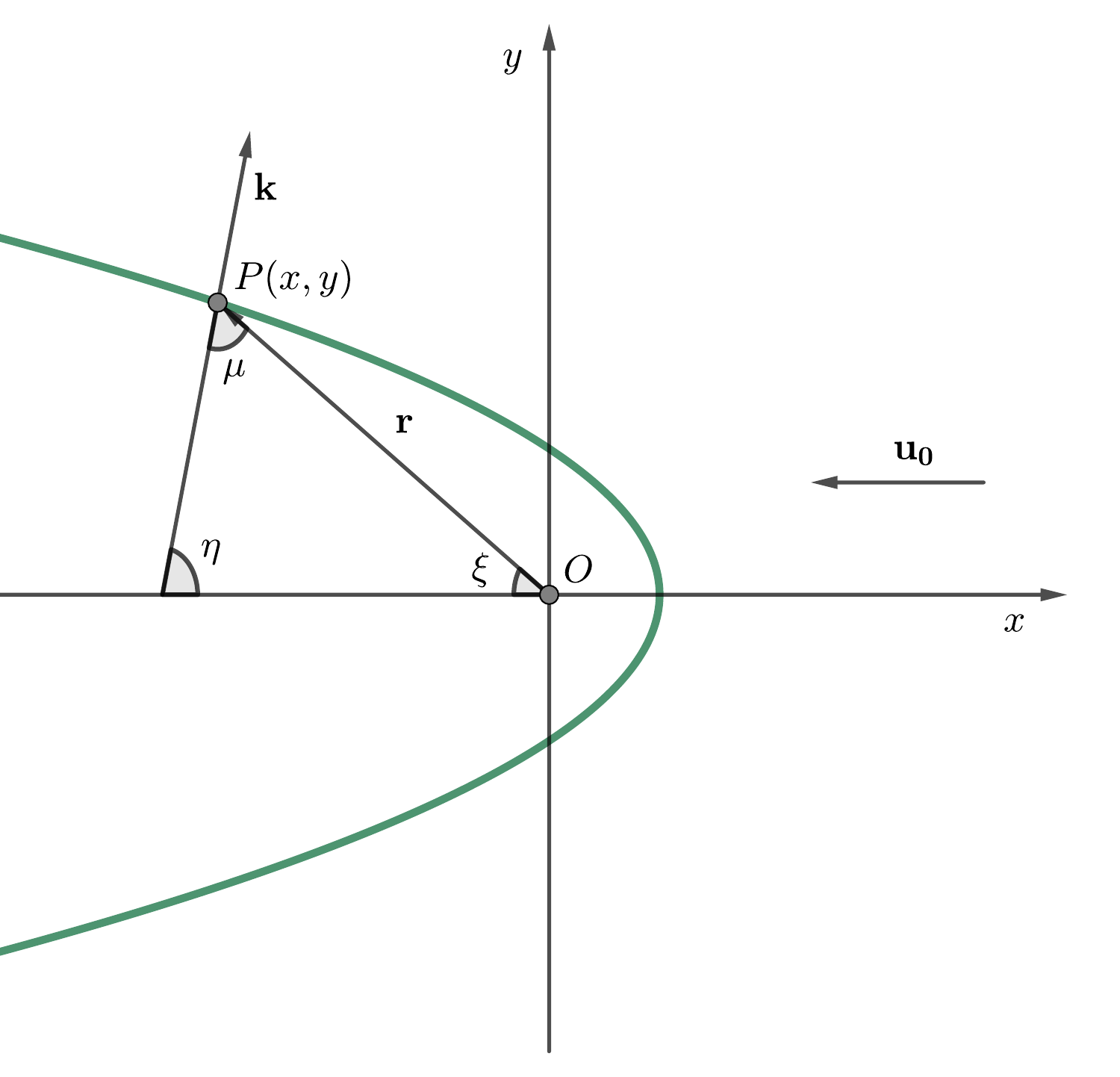}
    \caption{Geometry of the stationary wave crest of linear radiation outside the Mach cone. The group velocity $\nabla_{\bk}\omega(\bk)$ of the wave is compensated by the flow velocity $\bu$. The angle $\eta$ determines the direction of the wave vector $\bk$ and $\textbf{r}$ is the radius-vector of the point $P(x,y)$.}
    \label{fig1}
\end{figure}

It is clear that the stationary wave pattern can only be generated by a supersonic flow $M>1$ (or $u_0> c_0$), when the corresponding component of the flow velocity can compensate for wave propagation. A typical wave crest is shown in Fig.~\ref{fig1} for the flow incoming from the right. In a stationary wave generated by the flow past an obstacle, the wave vector is equal to the gradient of the phase $\theta$ of the wave, whereas the frequency must be equal to zero:	
\begin{equation} \label{eq9}
    \mathbf{k} = \nabla \theta,  \quad  G(k_x, k_y) = \omega = 0.
\end{equation}
Consequently, the equality of the cross-derivatives gives the equation	
\begin{equation} \label{eq10}
    \frac{\partial k_y}{\partial x} = \frac{\partial k_x}{\partial y}.
\end{equation}
The function $k_x = f(k_y)$ is defined in implicit form by the second Eq.~(\ref{eq9}), so we can rewrite Eq.~(\ref{eq10}) as an equation for $k_y$:
\begin{equation} \label{eq11}
\frac{\partial k_y}{\partial x} - f'(k_y) \frac{\partial k_y}{\partial y} = 0.
\end{equation}
It is known as the Hopf equation and its theory is well developed (see, e.g., Ref.~\cite{whitham-75}). It is clear that $k_x$ and $k_y$ are constant along characteristic curves which satisfy the equation
\begin{equation} \label{eq12}
    \frac{d y}{d x} = -f'(k_y).
\end{equation}
We are interested in the wave pattern far enough from a small obstacle, so we can take an obvious self-similar solution of this equation with characteristics stemming from the origin,
\begin{equation}\label{eq13}
    \frac{y}{x} = -f'(k_y).
\end{equation}
Then the definition of the derivative of the function defined implicitly yields
\begin{equation}
    \frac{y}{x} = \frac{\partial G / \partial k_y}{\partial G / \partial k_x}.
\end{equation}
At last, according to the definition of the angles in Fig.~\ref{fig1}, we have $y/x = \tan{\xi}$, and then straightforward algebra (see Ref.~\cite{gegk-07}) yields
\begin{equation}
	\tan{\xi} = \frac{\left( 2M^2 - 1 - \tan^2\eta \right) \tan\eta}{\left( M^2 + 1 \right)\tan^2\eta - \left( M^2 - 1 \right)}.
\end{equation}
Consequently, Eq.~(\ref{eq9}) gives the expression
\begin{equation}\label{eq16b}
  k=2\sqrt{M^2\cos^2\eta-1}
\end{equation}
for the wave number.

Another convenient parameter $\mu$ (see Fig.~\ref{fig1}) can be found from the relation
\begin{equation} \label{eq16}
    \tan\mu = 
    \frac{2M^2}{k^2} \sin2\eta.
\end{equation}
where $\xi + \mu + \eta = \pi$. The line of a constant phase can be obtained by integration the wave vector \eqref{eq9},
\begin{equation}\label{eq18}
    \theta = \int_0^r \mathbf{k} \cdot d\mathbf{r}.
\end{equation}
Since the wave vector is a gradient of the phase, the path of the integration can be chosen arbitrarily, and it is convenient to integrate along the line with $xi=\mathrm{const}$, so we get
\begin{equation}
    \theta = k r \cos\mu,
\end{equation}
where $r$ is the modulus of the radius vector. Then with the use of Eqs.~(\ref{eq16b}) and (\ref{eq16}) we find that the modulus of the radius vector can be expressed as a function of $\eta$,
\begin{equation}\label{eq19}
    r = \frac{4 \theta}{k^3} \sqrt{M^2(M^2 - 2)\cos^2\eta + 1}.
\end{equation}
At last, we obtain a parametric form of the line of a constant phase as
\begin{align}
    x &= -r \cos\xi = -\frac{4\theta}{k^3}\left( 1 - M^2\cos2\eta \right) \cos\eta, \nonumber \\
    y &= r \sin\xi = \frac{4\theta}{k^3}\left( 2M^2\cos^2\eta - 1 \right) \sin \eta.
\end{align}	
	
To obtain the equation for wave crests in notation of the LHY equation, 
we performed the inverse of the transformation \eqref{eq7}. According to the definition in Eq.~\eqref{eq18} we find that the transformation of the phase satisfies the relation $\theta \rightarrow\theta/c_0$, so 
\begin{align} \label{eq21}
    x &= \frac{4  c_0^2 \theta}{k^3 } \left[ M^2\cos2\eta - 1 \right] \cos\eta, \nonumber \\
    y &= \frac{4  c_0^2 \theta }{k^3}\left[ 2M^2\cos^22\eta - 1 \right] \sin\eta.
\end{align}	
In the last equation, we maintain the Mach number $M$ to simplify the notation. The angle $\eta$ varies in the interval
\begin{equation}
 -\arccos \left( \frac{1}{M} \right) \leq \eta \leq \arccos \left( \frac{1}{M} \right).
\end{equation}
These formulas show that this linear ``ship-wave'' pattern is located outside the Mach cone
which corresponds to the lines with $\eta=\pm\arccos(1/M)$.
	
	
\section{Oblique solitons in LHY fluid flow past an obstacle}
	
\begin{figure}[t]
	\centering
	\includegraphics[width=6cm]{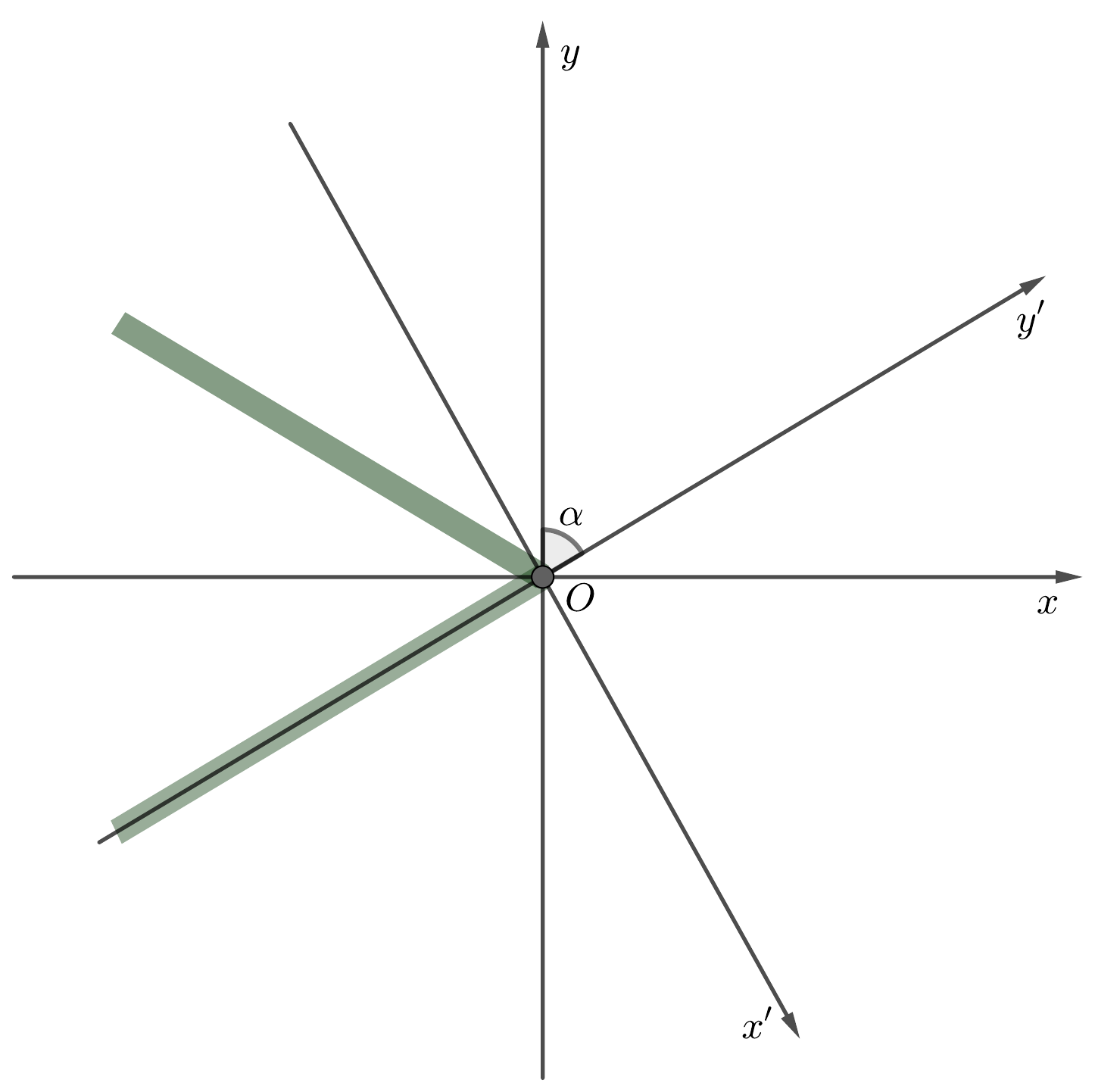}
	\caption{Geometry of dark oblique solitons inside the Mach cone.}
	\label{fig2}
\end{figure}

Another wave pattern observed within the Mach cone consists of two oblique dark solitons located symmetrically with respect to the $x$-axis, as is shown in Fig.~\ref{fig2}. In this section, we propose an analytical theory of this wave pattern. It is clear that these oblique solitons are developed from the ``shadow'' behind the obstacle and they attain a stationary form far enough from the obstacle. Therefore, we first find this stationary form assuming infinite length of a soliton. Such a soliton can be considered as a 1D solution of the generalized nonlinear Schr\"{o}dinger equation, so that its velocity is directed normally to the soliton's dip.
	
\subsection{DARK SOLITONS IN A 1D LHY FLUID}

A detailed discussion of the soliton solution and its stability can be found within the general theoretical framework developed in Refs.~\cite{kamchatnov-24, ks-09, bapa93}. The 1D LHY equation has the form
\begin{equation}\label{eq23}
    i \frac{\prt \psi}{\prt t} = -\frac{1}{2}\frac{\prt^2 \psi}{\prt \tilde{x}^2} + |\psi|^3 \psi, 	
\end{equation}
where it is assumed that $\tilde{x}$ is a coordinate normal to the soliton. We suppose for simplicity that the soliton moves with constant velocity $V$ through the condensate at rest. We make again the Madelung transformation
\begin{equation} \label{eq24}
    \psi(\tilde{x}, t) = \sqrt{\rho(\tilde{x}, t)}e^{i\int \tilde{u}(\tilde{x})\,d\tilde{x}},
\end{equation}
and obtain the hydrodynamic-like system
\begin{align} \label{eq25}
    \frac{\partial \rho}{\partial t} + \frac{\prt (u \rho)}{\prt \tilde{x}}  = 0,\nonumber \\
    \frac{\partial u}{\partial t} + u \frac{\prt u}{\prt \tilde{x}} + \frac{3}{2}\sqrt{\rho} 
    \frac{\prt \rho}{\prt \tilde{x}} =  \frac{\prt }{\prt \tilde{x}}  
    \left[ \frac{1}{4\rho}\frac{\prt^2 \rho}{\prt \tilde{x}^2} - 
    \frac{1}{8\rho^2}\left(\frac{\prt \rho}{\prt \tilde{x}} \right)^2 \right].
\end{align}
The standard assumption that the variables $\rho$ and $u$ only depend on the variable $\xi = \tilde{x} - Vt$ transforms this system to two ordinary differential equations which must satisfy the boundary condition 
\begin{equation} \label{eq26}
    u \rightarrow u_0, \quad \rho \rightarrow \rho_0\quad \text{ for }\quad |\xi| \rightarrow \infty,
\end{equation}
that is the soliton excitation is localized and the condensate is at rest far enough from the soliton.
	
Integration of the first equation of the system~\eqref{eq25} gives at once
\begin{equation}\label{eq27}
    u(\rho) = V \left( 1 - \frac{\rho_0}{\rho} \right).
\end{equation}
Then the second equation yields
\begin{equation} \label{eq28}
    \frac{1}{4\rho}\frac{d^2 \rho}{d \xi^2} - \frac{1}{8\rho^2}\left(\frac{d \rho}{d \xi}\right)^2 = 
    \rho^{\frac{3}{2}} - \rho_0^{\frac{3}{2}} + \frac{V^2}{2}\left( \frac{\rho_0^2}{\rho^2}  -  1 \right).
\end{equation}
This equation can be transformed with the use of the identity 
\begin{equation}\label{eq29}
    \frac{1}{4\rho}\frac{d^2 \rho}{d \xi^2} - \frac{1}{8\rho^2}\left(\frac{d \rho}{d \xi}\right)^2 = 
    \frac{1}{2\sqrt{\rho}}\frac{d^2}{d\xi^2} \left(\sqrt{\rho}\right),
\end{equation}
so that, after multiplication of Eq.\eqref{eq28} by $d (\sqrt{\rho}) / d\xi$ and easy integration, we get
\begin{equation}\label{eq30}
	\frac{d \rho}{d \xi} = \pm 2 \sqrt{R(\rho)},
\end{equation}
where
\begin{equation} \label{eq31}
	R(\rho) = 2\rho \left( \frac{3}{5} \rho_0^{\frac{5}{2}} - \rho_0^{\frac{3}{2}}\rho  + 
\frac{2}{5}\rho^{\frac{5}{2}} \right) - V^2\left( \rho_0 - \rho \right)^2.
\end{equation}
The dark soliton solution corresponds to $\rho$ changing between the two roots of $R(\rho)$, where $R(\rho)\geq0$, so these roots represent the maximal $\rho_0$ and minimal $\rho_m$ values of the density, respectively. In the center of the soliton we have ($\xi \rightarrow \xi_0$) $\rho \rightarrow \rho_m$, so integration of Eq.~\eqref{eq30} gives us the dependence $\rho=\rho(\xi)$ in implicit form,
\begin{equation} \label{eq.31}
    \xi (\rho) = \xi_0 \pm \frac{1}{2}\int_{\rho_m}^{\rho} \frac{d\rho}{\sqrt{R(\rho)}}.
\end{equation}
The relationship between the velocity $V$ of the soliton and the minimal value of the density $\rho_m$ at the soliton's center can be found directly from the equation $R(\rho_m) = 0$,
\begin{equation}\label{eq.32}
    V^2 = \frac{2\rho_m \left( \frac{3}{5} \rho_0^{\frac{5}{2}} + \frac{2}{5} 
    \rho_m^{\frac{5}{2}} - \rho_0^{\frac{3}{2}} \rho_m \right)}{(\rho_m - \rho_0)^2}.
\end{equation}

\subsection{OBLIQUE DARK SOLITON IN A QUASI-2D SYSTEM}
	
To extend our theory to the quasi-2D oblique dark soliton, we change the reference frame to the one where the obstacle is located at the origin of the coordinate system and the condensate flows from the right with the flow velocity $u_0<0$. We suppose that the soliton's dip lies along $y'$-axis rotated by the angle $\alpha$ with respect to the first reference frame,
\begin{equation} \label{eq33}
    x' = x \cos\alpha - y\sin\alpha \text{, \quad and \quad } y' = x \sin\alpha - y\cos\alpha.
\end{equation}
In this rotated coordinate system the flow velocity has the components $u'_0 = (u_0 \cos \alpha, u_0 \sin \alpha)$, and the soliton's velocity is compensated by the normal component of the flow velocity, so the soliton becomes stationary. 
This means that $V = -u_0\cos\alpha$ and we get the oblique soliton solution in the form
\begin{equation}
	x \cos\alpha - y\sin\alpha = \pm \frac{1}{2}\int_{\rho_m}^{\rho} \frac{d\rho}{\sqrt{R(\rho)}},
\end{equation}
where we took into account that the soliton stems from the origin where the obstacle is located, so $\xi_0=0$. Then, for a given value of $x$, the soliton's profile in $y$-direction is given by the formula
\begin{equation} \label{eq37}
    y(\rho, x)  =  x \cot \alpha \pm \frac{1}{2\sin \alpha}\int_{\rho_m}^{\rho} \frac{d\rho}{\sqrt{R(\rho)}}.
\end{equation}
The angle $\alpha$ is influenced by several parameters, including the flow velocity and the obstacle’s size and geometry. Its calculation lies beyond the predictive scope of our analytical approach. In our study, $\alpha$ was determined numerically using the relation $\alpha = \arctan(x_0/y_0)$, where $x_0$ and $y_0$ denote the coordinates of some point in the center of the soliton line.
	
\section{Numerical experiment and comparison with analytical results}
	
This section is devoted to comparison of analytical theory with numerical simulations.
	
\subsection{NUMERICAL METHOD}
	
The numerical approach employed in our simulations involves three transformations of the governing equation~\eqref{eq:LHY_adim}, similar to the method used in Refs.~\cite{kg-13}. The first transformation, applied to the homogeneous system, is used to fix the boundary conditions as \(\Psi = \Psi_0 e^{-i\rho_0 t}\). Next, we introduce the obstacle potential term, \(U_{\text{obs}} = U(x - vt, y)\), and perform a change of reference frame:
\begin{align*}
    &x' = x - vt; \quad y' = y; \quad t' = t; \\
    &\phi(x', y', t') = \psi(x, y, t) \, e^{-\frac{i v}{2} \qty(x - \frac{1}{2} v t)}.
\end{align*}

Finally, we apply a transformation to recover fixed boundary conditions:
\[
\varphi = \phi \, e^{\frac{i v}{2} \qty(x' + \frac{1}{2} v t')}.
\]
This leads to the equation derived in~\cite{hakim-97}:
\begin{equation}\label{eq:46}
\begin{split}
    i\pdv{\varphi}{t'} = & -\frac{1}{2} \qty(\frac{\partial^2 \varphi}{\partial x'^2} +\frac{\partial^2 \varphi}{\partial y'^2} )+ iv \pdv{\varphi}{x'} 
    -\rho_0 \varphi\\
    &+ U_{obs}(x',y')\varphi +\abs{\varphi}^3 \varphi,
    \end{split}
\end{equation}
which is solved numerically. 

Given the two-dimensional nature of the system, we use the Peaceman–Rachford (PR) method~\cite{quinney87} to handle spatial derivatives, combined with the split-step operator method for the ordinary terms. The PR scheme is well known for being unconditionally stable when applied exclusively to the dispersive components of the equation. In our implementation, it proved effective for evolving the system in time while maintaining numerical accuracy and stability.

To initialize the simulation, the wavefunction at $t = 0$ is set to a constant value, $\psi(x, y, 0) = 1$. The obstacle is modeled as an impenetrable circular barrier, incorporated through a piecewise-defined potential. Specifically, we define the potential $U(x, y)$ to be zero outside a circular region of radius $R_0$, and set it to a large constant $V_0 = 10^{12}$ within the region $\sqrt{x_i^2 + y_j^2} \leq R_0$, effectively creating a hard-wall condition that enforces $\psi = 0$ inside the obstacle. The typical spatial and temporal discretizations used in our simulations are $\Delta x = \Delta y = 0.20$ and $\Delta t = 0.001$, respectively.

\begin{figure}[t]
    \centering
    \includegraphics[width=8cm]{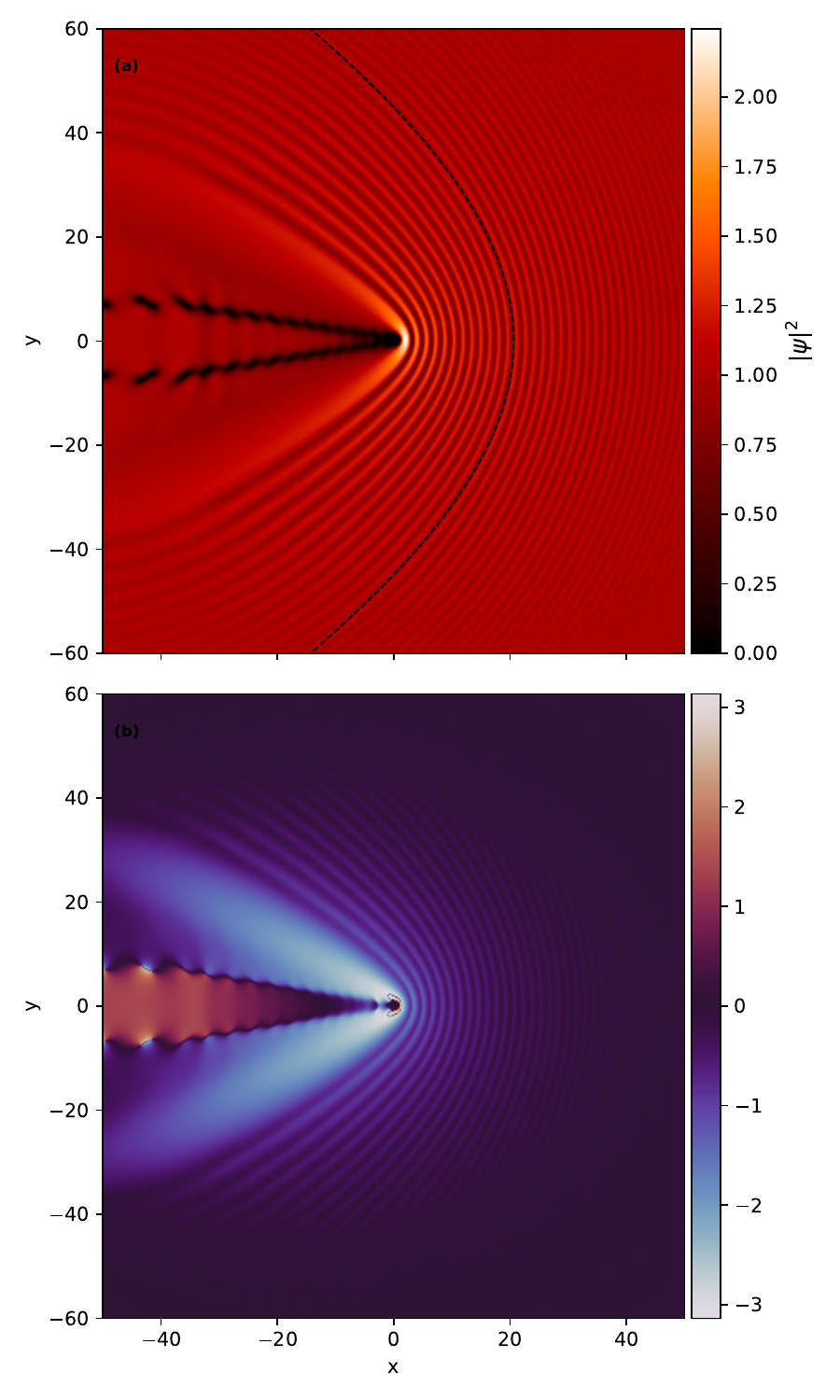}
    \caption{ (a) Comparison between the analytically predicted wave crest position [ Eq.~\eqref{eq21}] and numerical simulations Eq.~\eqref{eq:46}]. The analytical result is shown as a dashed line. (b) Numerical solution for the phase. The simulation was performed in a quasi-2D system with an impenetrable obstacle of radius $R = 1$, at evolution time $t = 25$ and Mach number $M = 2$.}
    \label{fig3}
\end{figure}
	
\subsection{LINEAR RADIATION}
	
\begin{figure}[t]
	\centering
	\includegraphics[width=8cm]{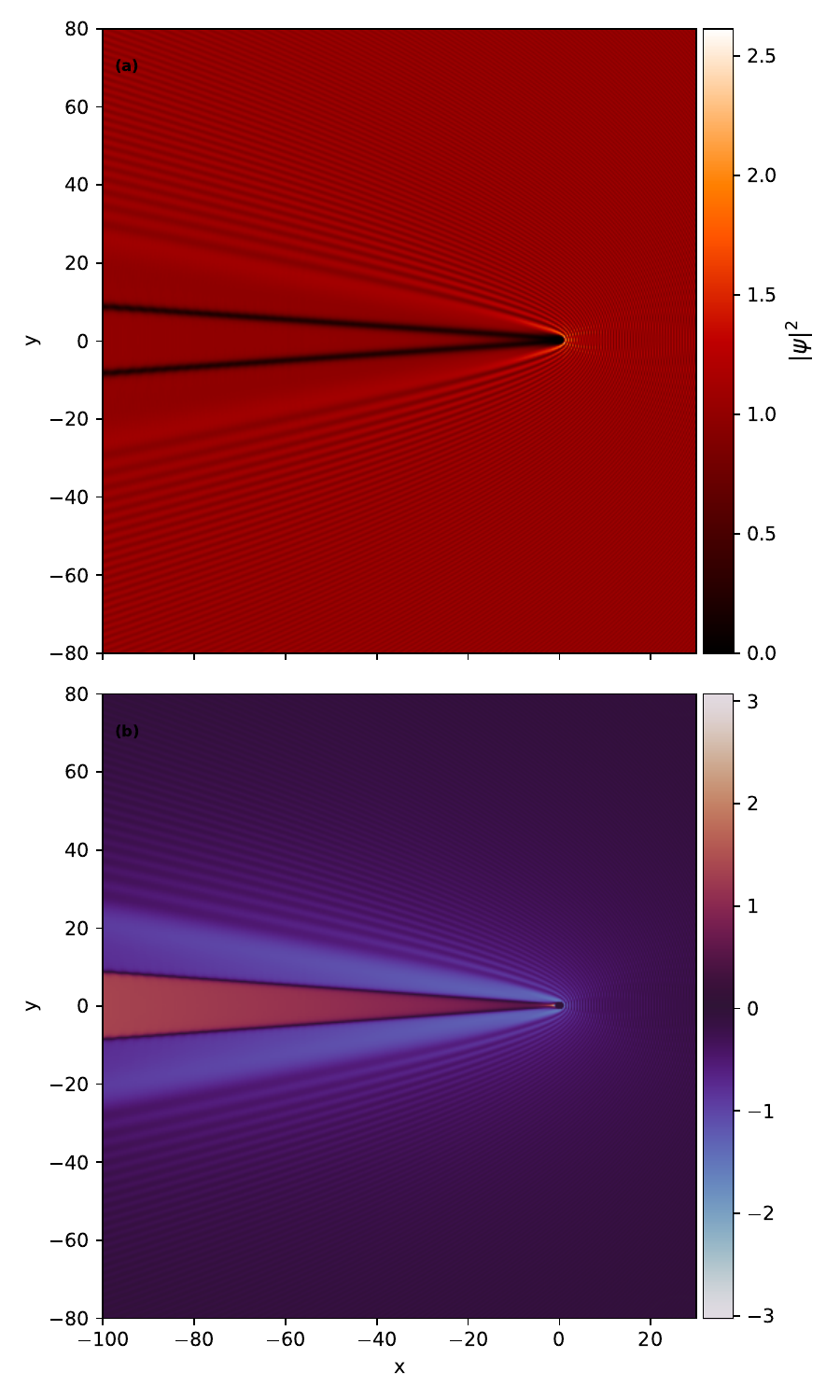}
	\caption{ Numerical simulation performed for a uniform quasi-2D system with an impenetrable obstacle of radius $R = 1$, evolution time $t = 25$, and Mach number $M = 5$. The density profile is shown in (a), and the corresponding phase in (b).}
	\label{fig4}
\end{figure}

As is shown in Fig.~\ref{fig3},  wave crests of linear radiation can be calculated analytically with good precision 
from Eq.~\eqref{eq21}. Moreover, Eq.~\eqref{eq21} predicts that just ahead of the obstacle, in the limit $y \rightarrow 0$, 
the wavelength vanishes as the flow velocity $|u_0|$ increases. This becomes clear by assuming $\eta \rightarrow 0$ 
and expanding Eq.~\eqref{eq21} in a Taylor series. This calculation results in the system:
\begin{equation}
	\begin{split}
		& x \approx \frac{\theta}{2\sqrt{u_0^2 - c_0^2}} + \frac{\theta(c_0^2 - 5u_0^2)}{\left( u_0^2 - c_0^2 \right) ^{3/2}} \eta^2, \\
		& y \approx \frac{\theta \left( 2u_0^2 - c_0^2 \right) }{2 \left( u_0^2 - c_0^2 \right) ^{3/2}} \eta,
	\end{split}
\end{equation}
which can be rewritten as a function $x=x(y)$,
\begin{equation} \label{eq39}
	x\left( y \right) \approx \frac{\theta}{2\sqrt{u_0^2 - c_0^2}} + \frac{4\left( c_0^2 - 5u_0^2 \right) 
\left( u_0^2 - c_0^2 \right)^{3/2}}{\theta \left( 2u_0^2 - c_0^2 \right)^2} y^2.
\end{equation}
Finally, from Eq.~\eqref{eq18}, we find that $\theta \approx kx$. Substituting this into Eq.~\eqref{eq39}, we obtain, in the limit $y \rightarrow 0$, the wavelength
\begin{equation}
	L \approx \frac{\pi}{\sqrt{u_0^2 - c_0^2}},
\end{equation}
which decreases as a function of the velocity $u_0$. This ``shortening'' of the wavelength can be seen when we compare the wave patterns for $M=2$ in Fig.~\ref{fig3} with the other one for $M=5$ in Fig.~\ref{fig4}. If we increase $M$ even more, then we observe in numerical simulations a ``continuous'' density distribution in the region of for small $\eta$. This means that the wavelength becomes here smaller than the size of the obstacle, so the interference of generated harmonics becomes destructive and it does not lead to formation of a regular wake wave pattern.
	
\subsection{OBLIQUE DARK SOLITON AND ITS STABILITY}
	
The oblique dark solitons produced by the LHY fluid inside the Mach cone attains a stationary form only when the critical flow velocity is reached. Below this critical velocity, the soliton is unstable and, similar to the case of the Gross-Pitaevskii equation, it breaks into vortex-antivortex pairs due to the snake instability. A discussion of this instability in the context of the Gross-Pitaevskii equation can be found in Ref.~\cite{kp-08} and qualitatively similar behavior is expected for the case of LHY dynamics. 
An example of a non-stationary street of vortices instead of a stationary oblique soliton is shown in Fig.~\ref{fig3}. Above the critical velocity $M>M_{c}$, the oblique dark soliton is formed, as is shown in Fig.~\ref{fig4}. A rough estimate based on our numerical results suggests that the critical Mach number for soliton formation lies in the range $3 \lesssim 
M_{c} \lesssim 3.5$. 
This value is more than twice the critical Mach number for oblique soliton formation in the dimensionless Gross–Pitaevskii equation, $M_c^{\text{GPE}} \approx 1.46$, as calculated in Ref.~\cite{kp-08}. Below this critical flow velocity, the oblique solitons are absolutely unstable and even for supersonic velocities $1<M<M_c$ vortex streets are generated instead. For subsonic flow velocities above $M_1\approx\sqrt{2/11}\approx0.43$, as was shown in Ref.~\cite{fpr-92}, generation of vortex streets leads to loss of superfluidity. Usually, these vortex streets consist of vortex pairs with opposite polarities.  However, as was shown in Ref.~\cite{sss-10}, in case of small values of the obstacle radius $0.04< R<0.13$ there is a region of velocities $0.4\lesssim M\lesssim0.56$ where B\'{e}nard-von K\'{a}rm\'{a}n-like streets of vortex pairs with the same polarity are generated. Obviously, these regions of parameters are beyond the region of generation of oblique solitons which we are interested in.

\begin{figure}[t]
    \centering
    \includegraphics[width=8cm]{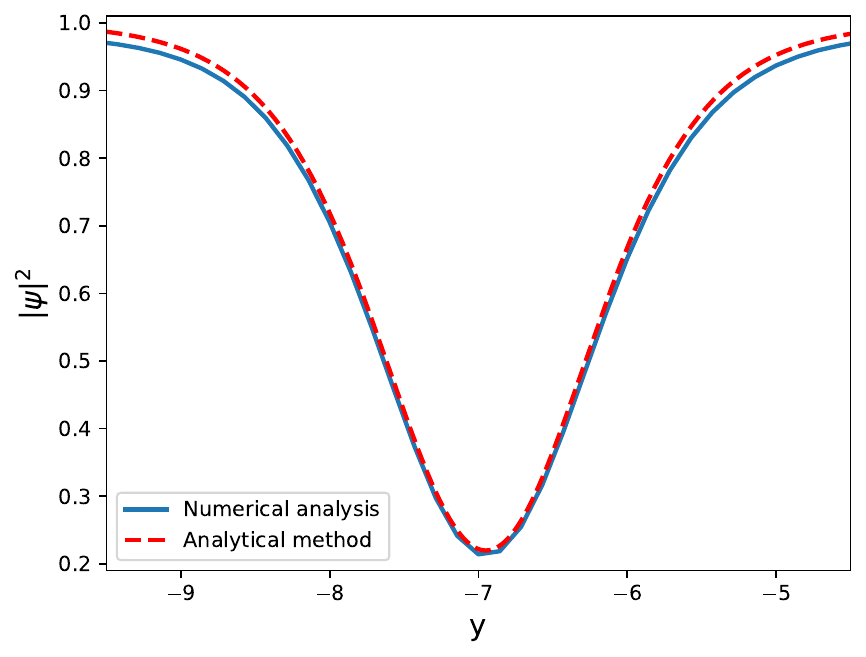}
    \caption{Analytical and numerical profiles of the dark soliton for $M = 5$ and $x = -80$. The analytical results were obtained using Eq.~(\ref{eq37}), where the value of the angle $\alpha=85.035^{\circ}$ was found from the numerical calculations.}
    \label{fig5}
\end{figure}

There is very good agreement between the analytical solution for the soliton density profile, given by Eq.~\eqref{eq37} for stable oblique solitons, and the numerical profile extracted from Fig.~\ref{fig4} at $x = -80$,  as demonstrated in Fig.~\ref{fig5}.

Using Eq.~\eqref{eq.32} and assuming $V = u_0 \cos{\alpha}$, we can express the relation between the soliton depth $\rho_m$ and the oblique soliton angle $\alpha$ as
\begin{equation}\label{eq50}
    \alpha(\rho_m) = \arccos\left( \frac{1}{|u_0|} \sqrt{ \frac{2\rho_m \left( \frac{3}{5} \rho_0^{5/2} + \frac{2}{5} \rho_m^{5/2} - \rho_0^{3/2} \rho_m \right)}{(\rho_m - \rho_0)^2} } \right).
\end{equation}

\begin{figure}[t]
    \centering
    \includegraphics[width=8cm]{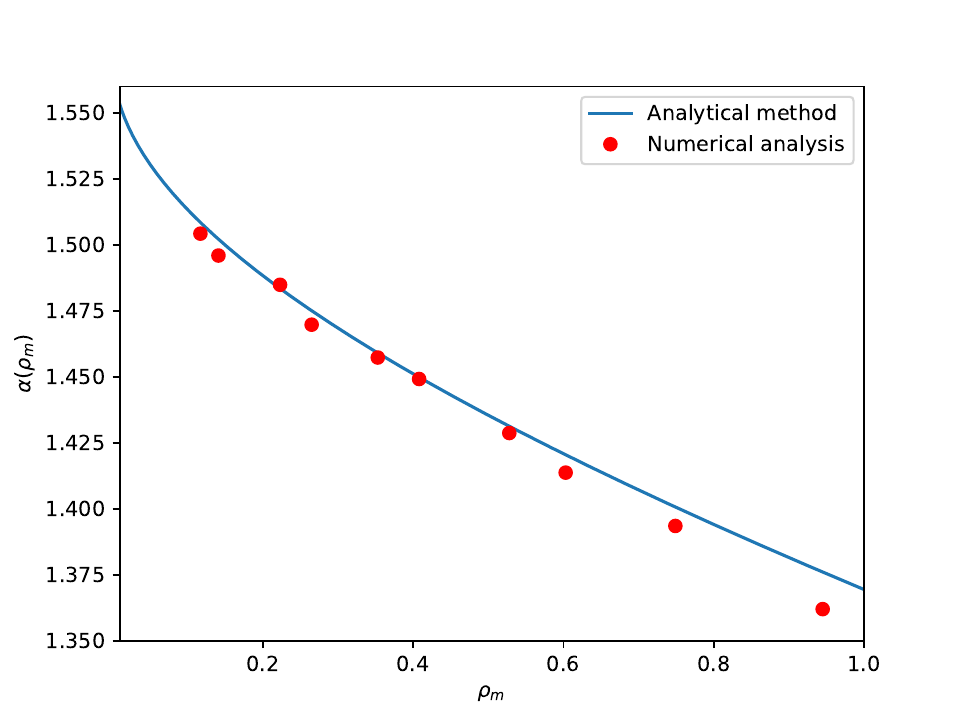}
    \caption{  Analytical  and numerical results for the minimum soliton density $\rho_m$ and the oblique soliton angle $\alpha$. The analytical results are given by Eq.~\eqref{eq50}. The numerical points are obtained by varying the obstacle radius $0.1 \leq R \leq 1.5$.}

    \label{fig6}
\end{figure}

This expression shows that although both the soliton depth $\rho_m$ and the oblique soliton angle $\alpha$ implicitly depend on the obstacle size, this dependence does not affect the accuracy of the theoretical prediction. This can be observed in Fig.~\ref{fig6}, where the pair $(\rho_m, \alpha)$ was obtained varying the value of the obstacle size in the range  $0.1 \leq R \leq 1.5$. However, for sufficiently large obstacles ($R \gg 1$), the number of oblique soliton pairs formed behind the obstacle increases, and the theory should then be applied to each soliton individually.

\section{Conclusion}
	
In this paper, we investigated analytically and numerically two types of wave excitations generated by a supersonic flow past an obstacle in a quasi-2D LHY quantum fluid. Numerical simulations confirmed that the wave crests of linear radiation, formed outside the Mach cone, are accurately captured by linear theory (see Fig.~\ref{fig3}). The structure of oblique dark solitons is well described by a quasi-one-dimensional analytical model (see Fig.~\ref{fig5}), demonstrating excellent agreement with the numerical results.

 As proposed in Ref.~\cite{lyu22}, the framework of a flow past a barrier serves as an experimental tool to measure and analyze critical velocities associated with the emission of collective excitations. The theoretical approach developed here should be useful to understand the behaviour of linear waves and solitons generated by the interaction of strong laser beams—approximated by an impenetrable barrier—with quasi-2D supersonic quantum droplets or strongly interacting mixtures, where LHY corrections dominate over the mean-field term.

As a direction for future work, we propose extending this approach to the case of supersonic flow past weak, penetrable obstacles, where the local flow velocity varies along the soliton line. As demonstrated in Ref.~\cite{ea-2007} for a cigar-shaped BEC, the properties of the obstacle also influence the behavior and stability of resulting excitations.

\begin{acknowledgments}	
L.F. Calazans de Brito acknowledges financial support from the Fundação de Amparo à Pesquisa do Estado de São Paulo (FAPESP), Grant No. 2023/17459-8. A.M. Kamchatnov was supported by the research project No. FFUU-2021-0003 of the Institute of Spectroscopy of the Russian Academy of Sciences (Sec.~III), and by the Russian Science Foundation (RSF), Grant No. 19-72-30028 (Sec.~IV). A. Gammal acknowledges support from FAPESP, Grant No. 2024/01533-7, and from the Conselho Nacional de Desenvolvimento Científico e Tecnológico (CNPq), Grant No. 306920/2018-2. G.H. dos Santos acknowledges support from the Coordenação Coordenação de Aperfeiçoamento de Pessoal de Nível Superior - Brasil (CAPES) - Finance Code 001.

\end{acknowledgments}
	
\bibliography{bib}
     
\end{document}